# QTC$_{3D}$: Extending the Qualitative Trajectory Calculus to Three Dimensions

Nikolaos Mavridis [a], Nicola Bellotto [b,*], Konstantinos Iliopoulos [a], Nico Van de Weghe [c]

[a] *Institute of Informatics and Telecommunication, NCSR Demokritos, Greece*
[b] *School of Computer Science, University of Lincoln, United Kingdom*
[c] *Department of Geography, Ghent University, Belgium*

**Abstract**

Spatial interactions between agents (humans, animals, or machines) carry information of high value to human or electronic observers. However, not all the information contained in a pair of continuous trajectories is important and thus the need for qualitative descriptions of interaction trajectories arises. The Qualitative Trajectory Calculus (QTC) (Van de Weghe, 2004) is a promising development towards this goal. Numerous variants of QTC have been proposed in the past and QTC has been applied towards analyzing various interaction domains. However, an inherent limitation of those QTC variations that deal with lateral movements is that they are limited to two-dimensional motion; therefore, complex three-dimensional interactions, such as those occurring between flying planes or birds, cannot be captured. Towards that purpose, in this paper QTC$_{3D}$ is presented: a novel qualitative trajectory calculus that can deal with full three-dimensional interactions. QTC$_{3D}$ is based on transformations of the Frenet-Serret frames accompanying the trajectories of the moving objects. Apart from the theoretical exposition, including definition and properties, as well as computational aspects, we also present an application of QTC$_{3D}$ towards modeling bird flight. Thus, the power of QTC is now extended to the full dimensionality of physical space, enabling succinct yet rich representations of spatial interactions between agents.

## 1. Introduction

As the epitome of the philosophy of Heraclitus (544-484BC) states: "*All entities move and nothing remains still*". Thus *change*, and especially *motion* (which is the primary sensory manifestation of change), are central elements in almost all philosophical-conceptual systems. One of the most important species of motion is *relative motion* between two entities, which forms an essential aspect of *spatial interaction*, for the case of objects construed as agents (humans, animals, or machines). Such spatial interactions between agents carry information of high value to *human observers*, as exemplified by the high-level interpretations and judgments that humans make when watching the Heider and Simmel movie (Heider & Simmel, 1944[1]), or by the rich semantic content of moving point abstractions of real-world events and everyday interaction scenes (e.g. reading gender from gait, Mather & Murdoch, 1994). Furthermore, such spatial interactions between agents carry invaluable information not only to human observers, but increasingly also to *electronic sensing systems*, for example those overlooking or assisting with crowd flows (Zhan et al., 2008), or traffic management (Buch et al., 2011). In recent years, geographical information scientists have intensively explored the relationships between multiple moving point objects. Research in this area has predominantly focused on the comparison of quantitative characteristics of trajectories such as azimuth, velocity, turning angle, acceleration, and sinuosity. An extensive overview is given in (Long & Nelson, 2013).

* Corresponding author.
 *E-mail address*: nbellotto@lincoln.ac.uk (N. Bellotto)

However, when observing the relative motion between two agents, not all the information contained in a pair of continuous trajectories is always important. For example, one might not really need the exact distance between two agents, but only the trend of change of relative distance or pose between them. Thus, the need for qualitative descriptions of interaction trajectories arises, abstracting unnecessarily complex complete quantitative representations. An adaptive representation of spatial trajectories of pairs or groups of objects, which can retain exactly as much qualitative information as needed for each application, can also be used for learning and reproducing interactive behaviors.

The Qualitative Trajectory Calculus (QTC), devised by (Van de Weghe, 2004), is a promising development towards this goal. A number of variants of QTC have been proposed in the past, including versions enabling the application of QTC to networks (Delafontaine et al., 2008), and shapes (Van de Weghe et al., 2005). However, an inherent limitation of the existing variations of QTC considering lateral movements (e.g. QTC Double Cross) is that they can only deal with two-dimensional motion. Therefore, complex three-dimensional interactions, such as those occurring between flying planes or birds, cannot be adequately captured. Towards such purpose, in this paper we propose QTC$_{3D}$: the first extension of QTC that can specifically deal with three-dimensional interactions.

Our representation is based on qualitative descriptions of transformations of the Frenet-Serret frames (Kreyszig, 1991) accompanying the trajectories of the moving objects. In more detail, the two Frenet-Serret frames corresponding to the two moving points consist of the tangent, normal, and binormal vectors. The relative motion between the two frames is modeled by the transformation that maps one frame to the other. Apart from the continuous model, the proper application of $QTC_{3D}$ in real-world sampled trajectories requires proper discretization, which is also devised and presented. Finally, an example towards qualitative modeling of the flight of a flock of birds is provided, illustrating the elegance and power of $QTC_{3D}$ for a compact representation of complex three-dimensional interactions while ignoring unnecessary detail and exposing only essential information.

---

[1] In this classic psychological experiment, a movie is shown to experimental subjects, where a set of simple geometrical figures (triangles, points, and lines) move in trajectories with respect to one another. However, when humans are asked to report what they have seen, they directly offer anthropocentric (or arguably, biocentric) interpretations of what they have seen: the triangles are reported as having affective state (angry, afraid, etc.), their relative motions are interpreted as intentional acts (chasing, confronting, hiding) and so on. All of this rich information is included not in the form of the figures, but just in the relative trajectories of them.

In this paper, we will proceed in section 2 by providing a discussion of relevant existing literature, followed in section 3 by a theoretical explanation including the definition of $QTC_{3D}$ and its fundamental properties. Then, in section 4, we will discuss computational aspects, and provide a version of $QTC_{3D}$ that can deal with discrete-time sampled trajectories. In section 5, we present an illustrative example of $QTC_{3D}$ towards modeling bird flight. Finally, we will close with a discussion, including future steps, followed by a conclusion. Overall, and most importantly, through this paper, the power of QTC will be extended to the full dimensionality of physical space, enabling succinct yet rich representations of spatial interactions between agents.

## 2. Background

Qualitative temporal and spatial reasoning about movement behavior has increasingly gained momentum over the last two decades, as scholars have begun to recognize the importance of qualitative reasoning in describing the common-sense background knowledge on which our human perspective on physical movements is based (Galton, 2000)(Guan & Duckham, 2011). In particular, various qualitative temporal calculi, such as the *Interval Calculus* (Allen, 1983) and the *Semi Interval Calculus* (Freksa, 1992), have been proposed. Along this line, a well-matured body of research has been developed regarding mereotopological relationships, as exemplified by the *RCC-calculus* (Randell et al, 1992) and the *9-intersection model* (Egenhofer & Herring, 1991).

Until recently however, there was a lack of academic work on calculi to represent trajectories of disjoint objects, hampering applications where most objects are disconnected, such as moving vehicles, pedestrians and animals. To address this shortcoming, (Van de Weghe, 2004) introduced the *Qualitative Trajectory Calculus* (*QTC*) to describe the relative motion of disconnected moving objects, providing an answer for many trajectory applications. As with other qualitative calculi, the theoretical framework of QTC has been thoroughly investigated by, among others, composition-tables (Van de Weghe, et al, 2006) and conceptual neighborhood diagrams (Van de Weghe & De Maeyer, 2005). This has been furthered by an implementation of QTC that is capable of describing real-world movements, both at time stamps (by QTCrelations) and during longer periods (by QTCanimations, being a sequence of QTCrelations) (Delafontaine et al, 2011). Such animations can represent all kinds of real-world interactions, including an overtake event (Van de Weghe et al, 2005a) and prey-predator interactions (Van de Weghe et al, 2005b).

Recently, QTC has been applied to analyze and implement human-robot spatial interactions. In the preliminary work of Bellotto (2012), a version of QTC dealing only with the linear distance between two agents (i.e. QTC Basic = $QTC_B$) was adopted to describe and implement simple spatial interactions, in which a robot and a human approached or moved away from each other. In (Hanheide et al., 2012), the human trajectory induced by a particular robot motion behavior in narrow spaces was analyzed using sequences of QTC states that included also lateral movements (i.e. QTC Double Cross = $QTC_C$). Combinations of $QTC_B$ and $QTC_C$ sequences were then exploited in (Bellotto et al., 2013) to design and implement human-robot spatial interactions with varying degrees of resolution, depending on the scenario and the desired robot's behavior. In all these cases, however, only 2D trajectories have been considered. The reason behind this is simple: in two dimensions, a unique line interconnecting the two moving points can be drawn, which divides the plane in two clearly defined regions. In three dimensions, a unique plane cannot be constructed between two points, and therefore no such clear partition exists.

Some previous work has considered qualitative spatial representations and reasoning on 3D regions (Albath et al., 2010). Also, an attempt has been made on the orientation of point objects, but only with respect to external reference systems (Pacheco et al., 2002; 2006). Furthermore, the complexity of the proposed models could limit their implementation and actual application to real-world problems. Thus, we need to resort to a novel constraint for QTC, in order to be able to capture the richness of interactions of a pair of three dimensional moving point objects.

## 3. Definition and Properties

a) A brief overview of $QTC_{2D}$

Let us start by providing a brief summary of the essentials of the traditional two-dimensional Qualitative Trajectory Calculus (Van de Weghe, 2004). The properties that $QTC_{2D}$ can retain are all the following ones, or specific subsets of them:

- A: Distance constraint for the first object, conventionally named *k*.
– means that it is approaching the second object, named *l*,
+ means that it is moving further away, and
0 means that its distance remains steady.

- B: Distance constraint, similar to A but with the objects *k* and *l* interchanged.

- C: Speed constraint; because of the dual nature we only need one such constraint.
– means that object *k* is slower than *l*,
+ means that *k* is faster than *l*, and
0 means that they move with the same speed.

- D: Side constraint for *k* with respect to vector *kl*:
− means that *k* is moving to the left of the line,
+ means that *k* is moving to the right of the line, and
0 means that it moves along the line.
- E: Side constraint, similar to D but with the roles of *k* and *l* interchanged.

- F: Angle constraint: define as $\theta_1$ the minimal angle between the velocity vector of *k* and vector *kl*, and $\theta_2$ the equivalent for *l*. Then we obtain
− if $\theta_1 < \theta_2$,
+ if $\theta_1 > \theta_2$, and
0 otherwise.

In order to help the readers better understand the above concepts, we provide the trajectories of two Moving Point Objects (MPOs) in Fig. 1 and the corresponding values of the constraints in table 1.

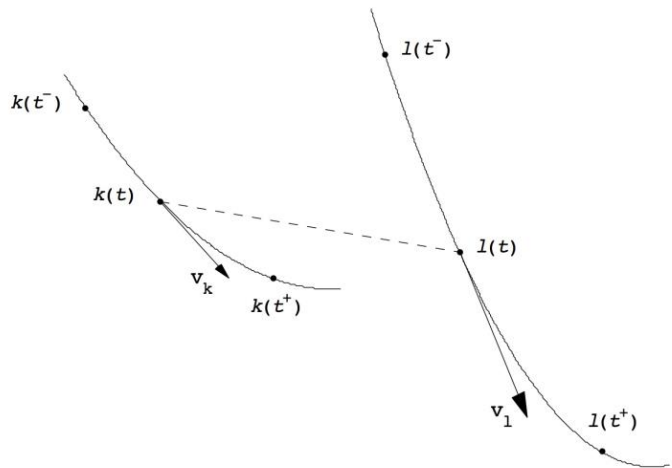

Fig. 1 – Trajectories of two MPOs

Table 1 – Constraints and their values for the MPOs of Fig. 1

| Constraint | Value | Explanation |
|---|---|---|
| A | - | *k* is moving towards *l* |
| B | + | *l* is moving away from *k* |
| C | - | *k* is slower than *l* |
| D | + | *k* is moving towards the right side of (*kl*) |
| E | - | *l* is moving towards the left side of (*lk*) |
| F | - | the angle between $v_k$ and (*kl*) is smaller than the angle between $v_l$ and (*lk*) |

By deciding to retain different subsets of the above constraints, we can obtain the following calculi, listed here in order of increasing complexity:
- $QTC_{B1}$: Supports relations A and B.
- $QTC_{B2}$: Supports relations A through C.
- $QTC_{C1}$: Supports relations A, B, D, and E.
- $QTC_{C2}$: Supports relations A through F.

For further explanation with respect to typical aspects of qualitative reasoning (e.g. dominance space, conceptual neighborhood diagrams, composition tables), we refer to (Van de Weghe, 2004).

b) Introducing $QTC_{3D}$

When extending QTC from 2D to 3D, analogous constraints to those outlined above have to be devised. Distance constraints (A, B), Speed constraint (C), and Angle constraint (F) can be easily generalized. However, as previously mentioned, there is no obvious analogue to the Side constraints (D, E).

The Frenet-Serret frame was thus chosen as our main instrument, as it provides a rich description of the kinetic properties of an object moving along a continuous and differentiable trajectory. The frame consists of three orthogonal vectors, which correspond to:
*t*: the unit vector tangent to the curve (eq. Ia & Ib),
*n*: the normal unit vector (eq. IIa & IIb), and
*b*: the binormal unit vector, i.e. a vector perpendicular to both *t* and *n* (eq. IIIa & IIIb).
The three vectors *t, n,* and *b*, create an orthonormal unit basis, thus attaching a frame of reference to each point in the trajectory (Fig. 2). Most importantly, this is a non-inertial frame, and one can furthermore prove that it is particularly well-behaved with regards to Euclidean motions, i.e. rotations and translations.

Therefore, the following definitions for $QTC_{3D}$ were chosen. Given two continuous three-dimensional trajectories $s_1(\tau)$ and $s_2(\tau)$, where $\tau$ is the continuous time variable belonging to $\mathbb{R}$:

STEP1) Calculate signs (-, 0, +) for all constraints A, B, C, and F as defined for $QTC_{2D}$ generalized from 2D to 3D

STEP2) Calculate the component vectors of the two Frenet-Serret frames, i.e. the tangents, normals, and bi-normals, as follows:

$t_1(\tau) = (ds_1/d\tau) / |ds_1/d\tau|$     (Ia)

$t_2(\tau) = (ds_2/d\tau) / |ds_2/d\tau|$     (Ib)

$n_1(\tau) = (dt_1/d\tau) / |dt_1/d\tau|$     (IIa)

$n_2(\tau) = (dt_2/d\tau) / |dt_2/d\tau|$     (IIb)

$b_1(\tau) = t_1(\tau) \times n_1(\tau)$     (IIIa)

$b_2(\tau) = t_2(\tau) \times n_2(\tau)$     (IIIb)

Now, our aim is to transform the frame $F_1(t_1, n_1, b_1)$ of the first moving object, to the frame $F_2(t_2, n_2, b_2)$ of the second moving object at the same time stamp. We thus need to find a transformation $T$, which transforms the first frame to the second, as follows:

$F_2 = TF_1 \Rightarrow T = F_2 F_1^{-1}$     (IV)

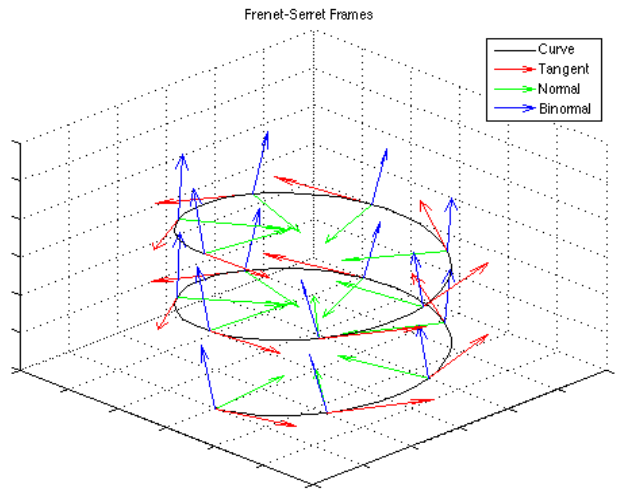

Fig. 2 - Illustration of the Frenet-Serret Frame

This transformation $T$, can be decomposed as the product of three rotations, which are usually known in the aeronautics literature as the yaw ($\phi$), pitch ($\theta$), and roll ($\phi$) (i.e. the so-called Tait-Bryan angles), as illustrated in Fig. 3.

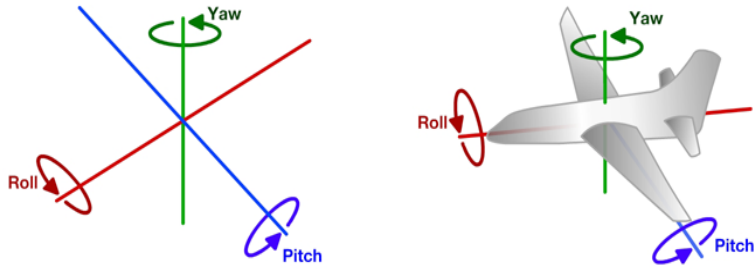

Fig. 3 – Yaw, Pitch, and Roll angles

We thus need to compute the three angles corresponding to the component rotations that multiply out to T, which we do through the following equations:

$$\text{assume } T = \begin{bmatrix} r_{11} & r_{12} & r_{13} \\ r_{21} & r_{22} & r_{23} \\ r_{31} & r_{32} & r_{33} \end{bmatrix}$$

$\psi = tan2(r_{21}, r_{11}), \psi \in [-\pi..\pi]$

$\theta = tan2\left(-r_{31}, \sqrt{r_{32}^2 + r_{33}^2}\right), \theta \in [-\pi..\pi]$

$\varphi = tan2(r_{32}, r_{33}), \varphi \in [-\pi..\pi]$

Then, in order to derive a meaningful qualitative representation for the quantitative representation of the three angles ($\phi$, $\theta$, $\varphi$), we need to quantize all possible values of this triplet to a set of qualitative (discrete) symbols, (-, 0,+) in QTC. For the ideal case of continuous trajectories (i.e. sampled with infinite uncountable sampling rate, and without corruption by measurement noise), we define the QTC symbols for each angle α in {$\phi$, $\theta$, $\varphi$}:

- If α = 0 → '0'
- If α < 0 → '-'
- If α > 0 → '+'

Thus, through this procedure, we derive the new QTC symbols G, H, I for the angles $\phi$, $\theta$, $\varphi$, respectively, which, in conjunction with the above A, B, C, and F, comprise the full QTC$_{3D}$ representation A, B, C, F, G, H, I.

## 4. From Ideal-Continuous Time to Real-Discrete Time

In order to apply the above in real-world time-sampled trajectories, one can use the Discrete Frenet-Serret Frame (DFF) (Hu et al., 2011). Here, equations (I)-(III) become, starting with tangent vectors:

$$t_1(\tau) = (x_1(\tau+1) - x_1(\tau)) / |x_1(\tau+1) - x_1(\tau)| \qquad \text{(Ia')}$$

$$t_2(\tau) = (x_2(\tau+1) - x_2(\tau)) / |x_2(\tau+1) - x_2(\tau)| \qquad \text{(Ib')}$$

We can then define[2] binormal vectors as:

$$b_1(\tau) = (t_1(\tau-1) \times t_1(\tau)) / |t_1(\tau-1) \times t_1(\tau)| \qquad \text{(IIa')}$$

$$b_2(\tau) = (t_2(\tau-1) \times t_2(\tau)) / |t_2(\tau-1) \times t_2(\tau)| \qquad \text{(IIb')}$$

and normal vectors as:

$$n_1(\tau) = b_1(\tau) \times t_1(\tau) \qquad \text{(IIIa')}$$

$$n_2(\tau) = b_2(\tau) \times t_2(\tau) \qquad \text{(IIIb')}$$

The discrete frames are:
$F_1(\tau) = (t_1(\tau), n_1(\tau), b_1(\tau)), F_2(\tau) = (t_2(\tau), n_2(\tau), b_3(\tau))$

The yaw, pitch, and roll angles are then calculated similarly to the continuous case. As can be seen in the equation below, for the quantization of continuous angle values to the three discrete symbols (-, 0, +), a threshold *Th* is used in this real-world case. This is required in order to delineate a symmetric band around the zero value of the angles, so that numerical deviations as well as measurement noise can be accounted for.

Thus, for $\alpha \in \{\phi, \theta, \varphi\}$, the mapping of values to symbols for the discrete case becomes:
- If α in [-*Th* .. *Th*] → '0'
- If α < -*Th* → '-'
- If α > *Th* → '+'

In this way, we are able to derive meaningful QTC$_{3D}$ symbol sequences from real-world sampled trajectories.

---

[2] Given that adjacent vectors are not parallel

## 5. A Real-World Example

In order to illustrate the utility of QTC$_{3D}$, we have chosen to apply it in a domain where rich 3D trajectories with complex interactions exist: bird flock flying. We utilize a micro-GPS derived dataset of pigeon flights from a recent paper published in Nature (Nagy et al., 2010). This dataset contains 4 homing- and 11 free-flights of at least 10 individuals each; in every flight, and especially in the homing ones, there exist a clear hierarchy of the roles of the pigeons. We then ask the following question: can information about pairs of interacting trajectories encoded in QTC$_{3D}$ be used towards distinguishing leader-follower bird pairs from other pairs? This is a typical interaction studied in reasoning about moving objects. In order to answer such a question, we have performed the following procedure.

First, we selected appropriate trajectory pairs (all of which were sampled at a temporal resolution of 200 ms), with and without Leader-Follower relations. As an example, we plot in Fig. 4 the trajectories of all pigeons of homing flight #1. Note that several pigeon trajectories have been truncated, effectively keeping only 2000 synchronized data points around the middle of the flight, in order to remove useless data before takeoff and after landing. For the case of Leader-Follower configurations, we would expect that a change in direction of the leader corresponds to a proportional change in the direction of the follower. That is, if pigeon $P_{Leader}$ moves towards a particular direction, then $P_{Follower}$ follows on a parallel direction after a short delay, which depends on the position of the pigeon within the flock hierarchy. In general, to compare these trajectories, one should consider this delay and temporally align the samples. However, in our case there is no need for relative time-shifting of the trajectories, given that the follower response has a delay smaller than the 200ms sampling interval.

Upon observation of the trajectories, we selected pigeon $P_H$ as the Leader. We can then classify the remaining pigeons in two categories, according to whether they closely follow the flight patterns of the leader or they significantly deviate from them:

a) Followers: pigeons $P_A$, $P_C$, $P_D$, $P_F$, $P_J$, $P_K$, $P_L$,
b) Non Followers: pigeons $P_G$, $P_I$.

We then extract the symbol distributions for all trajectory pairs. When we convert the trajectory pairs to QTC$_{3D}$ strings, they will consist of 7-tuples of (-,0,+). The important information for our task is contained in the sub-triplet {G, H, I} of the full QTC$_{3D}$ 7-tuple; after all, this is what differentiates QTC$_{3D}$ from QTC$_{2D}$. In this triplet there exist $3^3$=27 possible combinations of symbols. We try to estimate the probability distribution of these combinations by calculating a histogram based on their occurrences. Our ultimate goal in this section will be to differentiate between trajectory pairs of Leader-Follower and Leader-NonFollower roles: we will show this is possible using the ratio of entropies from the histograms of the QTC$_{3D}$ symbol distributions of Leader-Follower vs. Leader-NonFollower trajectories, while differentiation would not have been possible using the QTC$_{2D}$ symbols alone (i.e. without the new symbols {G, H, I}.

First of all, we need to make an informed choice of the appropriate thresholds for the derivation of QTC$_{3D}$. Towards that purpose, we will first investigate the histograms of the distributions of the Tait-Bryan angles. Fig. 5 and 6 display the histograms of the yaw, pitch, and roll angles, for the Leader-Follower and Leader-NonFollower respectively, bundled in bins of approximately 8 degrees each. We have chosen 8 degrees per bin for this visualization in order to have enough samples for each bin, so that the resulting curve is smooth and closer to the actual distribution.

In Fig. 5 and 6 (left) we see the frequency distribution of the yaw angles for the aforementioned case of homing flight #1, and we can already identify how discriminative it can be for the possible categories of pairs. If we set the threshold at 24 degrees, then the total probability mass created by the sum of the central 3 bins will map to the probability mass of the '0' symbol, while the bins on the right will map to the '+' symbol and the bins on the left will map to '-'. Note that, in the case of Leader-Follower, there will be a larger total mass for the '0' symbol, as the sum of the 3 central bins for the Leader-Follower case is larger than the sum of the equivalent ones for the Leader-NonFollower case. Correspondingly, the total mass for each of the '+' or '-' symbols will be smaller for the Leader-Follower distribution when compared to the Leader-NonFollower one. Thus, if we were taking the entropy of the single symbol corresponding to the yaw angle, the entropy of the Leader-Follower distribution would be smaller than the entropy of the Leader-NonFollower.

In practice, though, we will use all 3 angles (yaw, pitch, and roll), not individually but in conjunction in order to create the $3^3=27$ possible combinations of symbols, and we will take the entropy over this 27-symbol distribution (and **not** the 3 entropies of the three 3-symbol distribution corresponding to each angle separately). As we shall see, when we combine the symbols for all 3 angles, we will expect significantly different probability distributions. The key thing here is to choose an appropriate threshold *Th* to get a meaningful band of '0' symbols.

Because the Leader-Follower behavior requires the tracking of the direction of the flight of the leader by the follower, we expect that whenever this direction does not change, the follower will be aligned to it. This will happen not only in terms of direction, but also in terms of velocity and acceleration, if the alignment between leader and follower is to remain and the distance between the two is controlled by the follower with the goal of being kept constant. Thus, the two Frenet-Serret frames will be almost aligned for the period of time that the leader is not changing significantly his trajectory. In this case, the Tait-Bryan angles corresponding to the transformation needed to align one Frenet-Serret frame to the other will frequently have values close to zero. Therefore, the resulting distribution of the quantized QTC symbols corresponding to these angles will exhibit more triplets containing one or more '0's for the Leader-Follower case, as compared to the Leader-NonFollower one. In the latter case, the two Frenet-Serret frames will be generally more unrelated, and thus the transformation needed to map one to the other will be more random. In conclusion, we expect the distribution of QTC symbols for the yaw, pitch, and roll angles for the case of Leader-NonFollower to be closer to uniform (larger entropy) as compared to the symbol distribution for the Leader-Follower case (smaller entropy, given that the distribution is less uniform, with a larger percentage of triplets that contain '0's).

We then decided to investigate the entropies of the two QTC symbol distributions (i.e the symbols corresponding to the trajectory of the Leader-Follower pair, and the symbols corresponding to the Leader-NonFollower pair) and to use these entropies ratio as a discriminative feature for Leader-Follower vs. Leader-NonFollower pairs.

Those entropies were defined as per usual:

$$H(X) = -\sum_{i=1}^{n} p(x_i) \log_2 p(x_i) \tag{IVa}$$

And the ratios were defined as such:

$$ratio = \frac{\sum_{i=1}^{N} H_i}{\sum_{j=1}^{M} H_j} \cdot \frac{M}{N} \tag{IVb}$$

where we assume that we have N pairs of the type Leader-Follower and M pairs of the type Leader-NonFollower.

Indeed, our data indicated that for any appropriate choice of angle threshold *Th* equal to or above five degrees, this was the case: for example, for the trajectories displayed in Fig. 7, with a chosen angle threshold *Th* equal to 10 degrees, the Frenet-Serret transformation angle QTC symbol distribution entropy for Leader-Follower was 3.26, compared to 4.01 for the other case of Leader-NonFollower; this kind of relation in differences was found in all the other trajectory pairs we investigated. Therefore, the introduction of the novel symbols G, H, and I in QTC$_{3D}$, which accounts for the rotation angles required for matching the Frenet-Serret frames of the moving objects, was the catalyst towards providing us with clear discrimination between qualitatively different pairs of trajectories.

The question though arises: How well does this result generalize to other such pairs of trajectories? For that purpose, we have examined the other combinations that exist in our dataset, and we have found that the inequality still holds in all these cases. In more detail, the four homing flights contain a total of 27 Leader-Follower pairs and 7 Leader-NonFollower pairs. In Fig. 7 one can see the mean, mean+std, and mean-std entropies for each of the two classes of pairs. For a threshold *Th*≥ 10 degrees, we can see that the class corresponding to Leader-NonFollower consistently has a higher entropy when compared to the Leader-Follower class, hinting at the fact that this is statistically significant and therefore confirming our initial hypothesis that QTC$_{3D}$ can be used towards distinguishing Leader-Follower bird pairs from other pairs.

As a further and final elaboration of this result, the reader can check the bottom part of Fig. 7, which contains the ratio of the mean entropies for the two classes, which indicates that the entropy ratio is less than 1 for all cases of a non-trivial angle threshold, i.e. above 5 degrees. Thus, indeed, the entropy ratio criterion generalizes well, and the novel symbols G, H and I in QTC$_{3D}$, which were not part of the traditional QTC$_{2D}$, are indeed the catalyst towards this achievement, illustrating the power and applicability of QTC$_{3D}$.

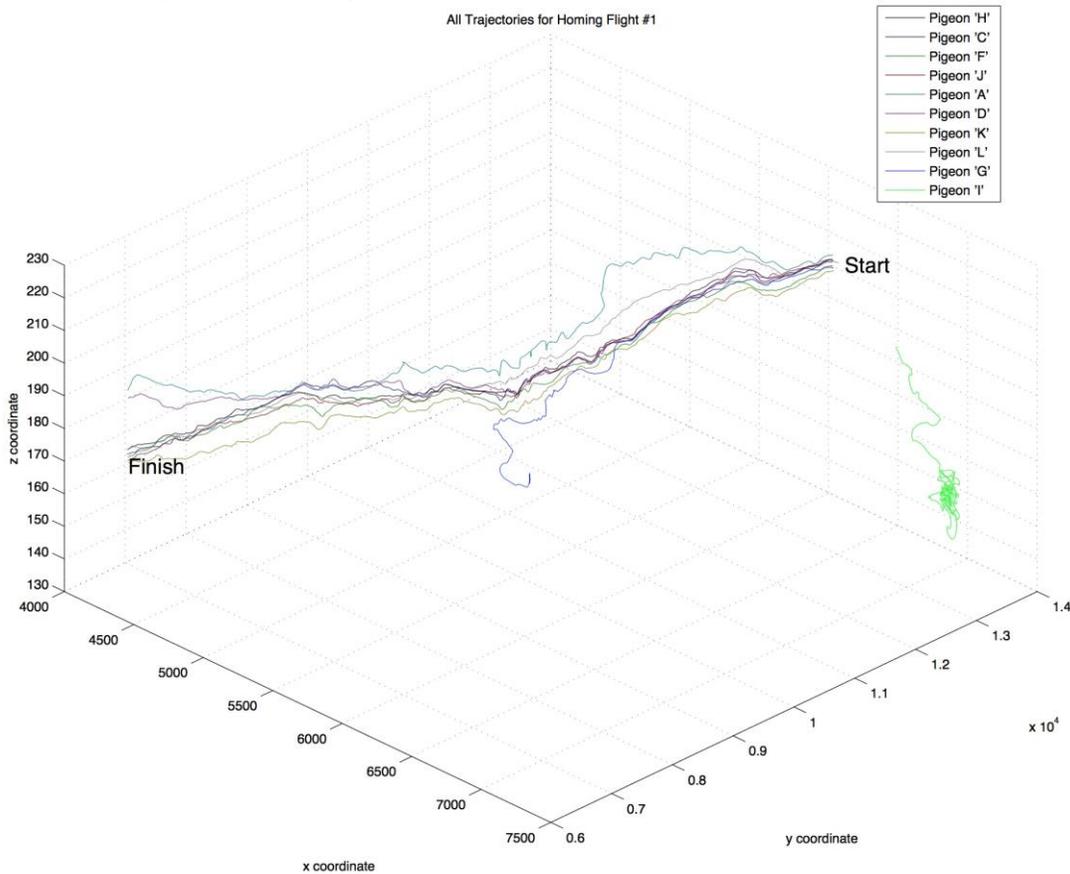

Fig. 4 – Truncated flight paths of all the pigeons of homing flight #1.

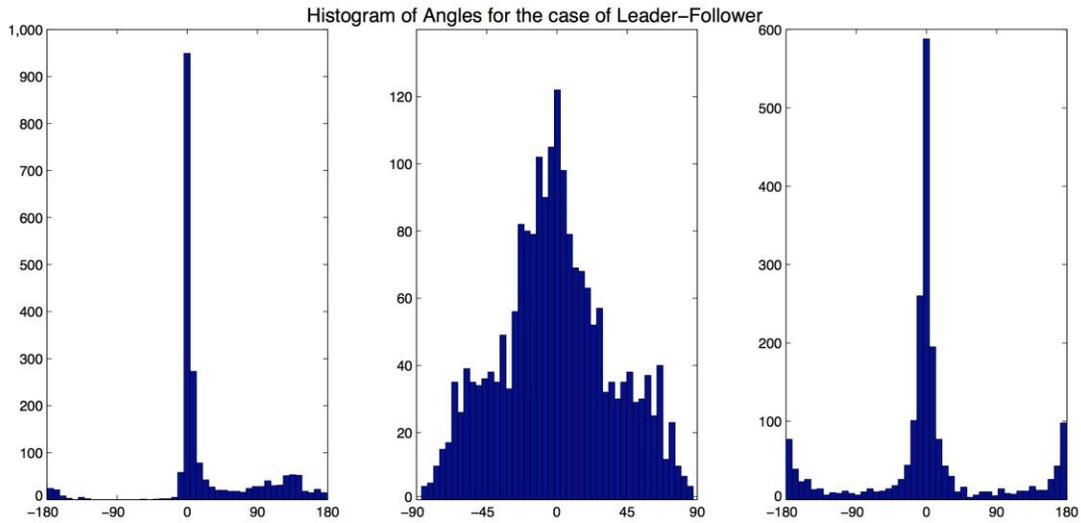
Fig. 5 –Yaw, Pitch, and Roll for the cases of Leader ('H') and a Follower ('A')

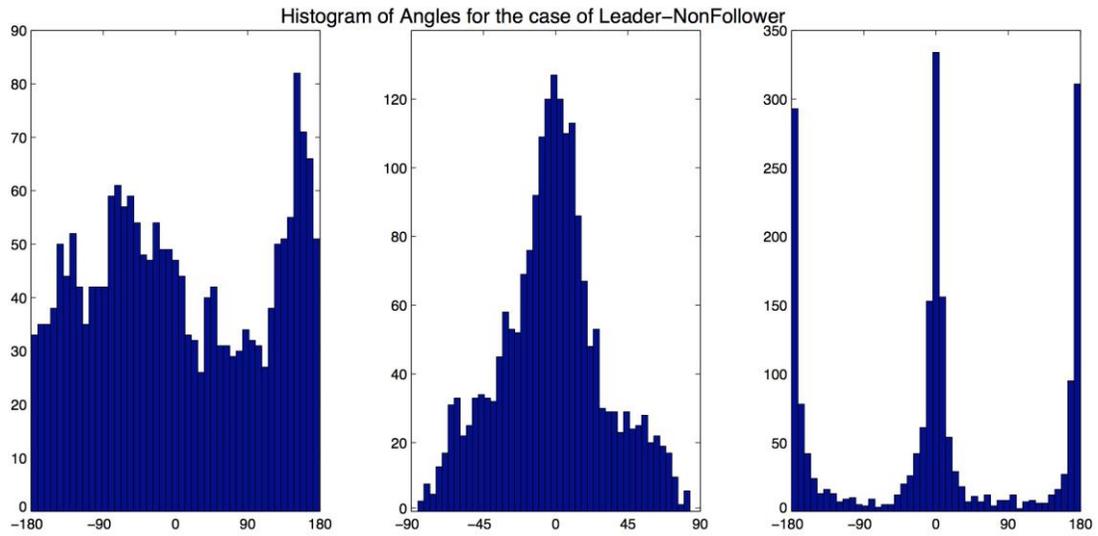
Fig. 6 – Yaw, Pitch, and Roll for the cases of Leader ('H') and a NonFollower ('I')

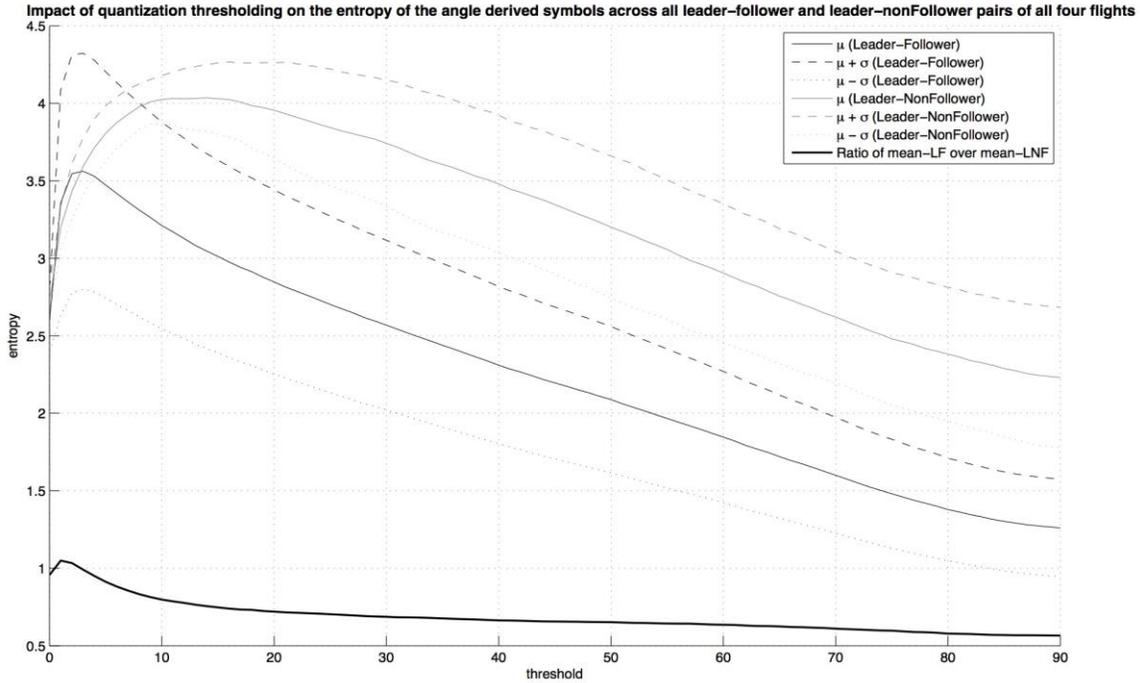

Fig. 7 - Quantization for the '0' symbols and impact on ratios, across all Leader-Follower and Leader-NonFollower pairs for all 4 homing flights, assuming that we only use properties G, H, and I. The slight rise in the beginning is easy to understand: before some meaningful quantization, we are almost completely stripped from '0' symbols, hence the smaller entropy. Once we account for that, however, the entropy quickly drops.

## 6. Discussion and Future Steps

Having introduced QTC$_{3D}$, and having illustrated its benefits through the bird flight scenario using real-world trajectories, let us now discuss an important point, which is concerned with the need for thresholding. In real world situations, most often apart from time sampling (discrete-time QTC) there is also noise in our trajectory measurements. The problem is that small perturbations in the positions of the MPOs may significantly affect the exported QTC symbols. As an example, consider the cases where two objects would be moving with the same speed. Clearly, even the slightest noise will cause change to the '0' symbol for the speed constraint to become either '+' or '-', and this is unacceptable. Thus, it is very important to define thresholds around zero: but how to set these thresholds? Note that, because of the nature of the equations and the calculations that they imply (Euclidean distances for the distance constraint, cross-products for the Side constraints etc.) it is not possible to define a meaningful universal threshold for all the QTC constraints.

If we can model the statistical behavior of the noise we are dealing with, we can attempt to fine-tune the thresholds accordingly (analytically or empirically). As a qualitative criterion for optimal tuning, one could try to minimize a reconstruction error, such as the symbol difference between a noise-free zero-threshold QTC sequence and the noisy thresholded version of the sequence. Alternatively, other application-specific criteria can be used for tuning the threshold, including for example variations of discriminability between sequences corresponding to different categories.

Regarding potential application scenarios, an obvious domain would be modeling of insects, airplanes, and unmanned aerial vehicles (UAVs) flight, or even fishes and unmanned underwater vehicles (UUVs). Furthermore, and quite importantly, $QTC_{3D}$ can be utilized not only towards the analysis of trajectories, as is the case in our bird flight example of the previous section, but also towards synthesis: i.e. given a specific QTC sequence, creating behavioral controllers for a robot or UAV/UUV that can perform the correct movements in response to a moving interaction partner, in order to satisfy the prescribed QTC sequence. An example of hand-crafted controller informed by QTC analysis and applied to Human-Robot Spatial Interaction can be found in (Bellotto et al., 2013). For the automated solution of the more general problem, which is the generation of prototypical trajectories of two objects satisfying a given QTC sequence, one needs to provide a solution to the so-called "Inverse QTC problem", which was for the first time provided in (Iliopoulos et al., 2014).

Other interesting application domains are the arts and sports. Group dance movements, for example, contain intricate yet often highly structured patterns of motion; QTC could be used not only towards analysis of human relative trajectories as moving point objects, but also by placing moving point objects at important human body points, and then describing the relative motions within a dancer's body or across dancer's body points (Chavoshi et al., 2014). Similar considerations can be done for sports analytics, where $QTC_{3D}$ could find extensive application, given the importance of the third dimension in this domain.

In terms of future steps, we are currently working not only with the theoretical formalization of thresholding techniques and generalization of the inverse QTC problem, but also with the practical application of QTC in various domains (e.g. robotics, sport, etc.), where a multitude of interesting extensions remain to be explored towards the efficient handling of multiple moving point objects, including groups and centers of symmetry of objects, opening up opportunities for widespread applications of $QTC_{3D}$.

## 7. Conclusion

Spatial interactions between natural or artificial agents (humans, animals, or machines) can be found almost everywhere, and carry information of high value to human or electronic observers. However, not all the information contained in a pair of continuous trajectories is important and thus the need arises for adaptive abstractions, such as qualitative descriptions of interaction trajectories.

In this paper we have presented $QTC_{3D}$, a novel qualitative trajectory calculus that can deal with full three-dimensional interactions, thus moving beyond the limitations of the traditional two-dimensional approach. QTC$_{3D}$ is based on transformations of the Frenet-Serret frames accompanying the trajectories of the moving objects. Apart from the theoretical exposition, including definition and properties, as well as computational aspects, we have also presented in detail a real-world application of QTC$_{3D}$ towards modeling bird flight, using real trajectories, illustrating the benefits of our approach. This opens up a wide range of real-world applications where such representation provides the catalyst for effective analysis and synthesis of complex spatial group behaviors.